\documentstyle{aipproc}
\input epsf

\def \beq{\begin{equation}}
\def \eeq{\end{equation}}
\def \beqn{\begin{eqnarray}}
\def \cn{Collaboration}

\def \eeqn{\end{eqnarray}}
\def \ite{{\it et al.}}

\def \suc{SU(4)$_{\rm C}$}
\def \sut{SU(4)$_{\rm T}$}

\def \v#1#2{V_{#1#2}}

\def \ubl{U(1)$_{\rm B-L}$}
\begin{document}

\title{Explorations of Compositeness\footnote{Enrico Fermi Institute report
EFI-98-60, December, 1998, hep-ph/9812537.  Based in part on an invited talk at
{\it Workshop on New Strong Dynamics for Run II of the Fermilab Tevatron},
Fermilab, Oct.~30--31, 1998.}} 

\author{Jonathan L. Rosner}
\address{Enrico Fermi Institute and Department of Physics \\
University of Chicago \\
5640 S. Ellis Avenue, Chicago IL 60637}

\maketitle

\begin{abstract}
Some of the motivations for quark and lepton compositeness, and some problems
associated with present schemes, are noted.  One model is discussed in
which quarks and leptons are taken as composites of spin-1/2 fermions $F$ with
charges $\pm 1/2$ and spinless bosons $\bar S$ with charges $1/6$ and $-1/2$.

\end{abstract}


\section{Introduction}

The breaking of electroweak symmetry is one of the unsolved problems of the
``Standard Model.''  It can be parametrized in terms of vacuum expectation
values of one or more Higgs boson fields, but there is undoubtedly a deeper
structure underlying the Higgs sector of the theory.  The Higgs field(s)
could be elementary but with mass(es) protected from large quartic
divergences by cancellations in loop diagrams among particles of different
spins.  Supersymmetry is a convenient scheme, possibly the only one, for
implementing this idea.  Alternatively, the Higgs field(s) could be composed
of more elementary objects, perhaps fermions and antifermions.  Such schemes
are collectively referred to as ``dynamical symmetry breaking,'' since they
involve a new strong dynamics to bind the constituents of the Higgs boson(s)
to one another.

The simplest incarnation of dynamical electroweak symmetry breaking
\cite{WTC,STC}, known as ``technicolor,'' envisions the condensation of
fermion-antifermion pairs under the influence of a QCD-like force, but one
which becomes strong at about 2650 times the QCD scale.  The corresponding
``techni-pions'' become the longitudinal components of $W^\pm$ and $Z^0$. 
(For reviews, see \cite{TCrevs}.)

Technicolor works adequately to induce gauge boson masses, but electroweak
symmetry breaking is also manifested in quark and lepton masses.  To explain
these, technicolor must be ``extended'' \cite{ETC}.  So far, no scheme of
extended technicolor has proved adequate to explain the pattern of quark and
lepton masses, the magnitudes and phases of elements of the
Cabibbo-Kobayashi-Maskawa (CKM) matrix, or the suppression of flavor-changing
neutral currents. 

If the Higgs bosons are composite, why not quarks and leptons as well? The
patterns of masses and electroweak transitions is reminiscent of a level
structure.  However, the difficulties in constructing light composite fermions
\cite{tH} make either supersymmetry or technicolor appealing by comparison. One
has no {\it a priori} reason to expect an almost pointlike object such as a
quark or lepton to have a mass much less than the characteristic scale
$\Lambda$ of its structure.  Present lower limits on $\Lambda$ in various
processes exceed one to several TeV.  The near-masslessness of the observed
fermions, on this scale, could be attributed to a nearly exact chiral symmetry,
realized in the Wigner-Weyl sense.  Using a criterion proposed by 't Hooft
\cite{tH}, one can test theories for this chiral symmetry by comparing gauge
anomalies as realized by composite fermions and by their constituents.  The
results should be the same.  The construction of realistic models satisfying
this anomaly-matching condition has proved extraordinarily difficult. 

What has happened in the 20 years since 't Hooft first proposed his condition?
A good review of the status of the early model-building efforts is given by
Peccei \cite{Peccei}. No other classes of theories have provided much insight
into the pattern of quark and lepton masses and mixings (though some partial
glimmers have emerged in the context of supersymmetry and string theory
\cite{Ramond}).  It may be time to re-examine the idea of compositeness as
``one more layer of the onion.''  In the past few years, some powerful
techniques have emerged for the study of light composite fermions \cite{Se}.
These are being exploited in model-building exercises by several groups (see,
e.g., \cite{comps}). 

The present talk is intended as a recapitulation of some previous efforts to
construct and test models of composite fermions, and an inducement to consider
some of the attractive features of quark and lepton compositeness.  The
compositeness of hadrons (made of quarks) seems beyond question now, even
though it may not have provided a full blueprint to properties of the proton.
For example, we still do not know how much of a proton's spin is carried by
quarks, but the classification of the proton and seven other spin-1/2 baryons
into an octet of flavor SU(3) \cite{GMN} is much more easily visualized in
terms of the quark model.  Similarly, we might hope to understand the apparent
existence of three families of quarks and leptons without solving all the
associated dynamical problems. 

In Section 2 we recall some reasons for expecting quarks and leptons to be
members of a level structure.  We note in Section 3 the suggestion that a heavy
top might be singled out as having a special role in dynamical electroweak
symmetry breaking, and discuss an Adler-Weisberger sum rule which hints that
the top quark may not be any more special than any other quark if the
electroweak scale is set by TeV-scale physics.  We outline in Section 4 the
motivation for regarding electroweak symmetry breaking as a scaled-up version
of QCD, and discuss an extension of such a scheme to the description of
composite quarks and leptons in Section 5. Some experimental signatures of this
specific model are mentioned in Section 6.  The model has features (Section 7)
which suggest that it might be made supersymmetric with relatively little
modification.  We discuss some of its open questions in Section 8.  Some
specific predictions of a model based on a superstrong SU(4) are noted in
Section 9. An alternative composite model, in which families are viewed as
distinct spin configurations of constituents, is discussed in Section 10, while
Section 11 summarizes. 

\section{Why expect a ``level structure''?}

The most compelling suggestion that quarks and leptons form some sort of level
structure is the pattern, illustrated in Fig.~1 \cite{DPF}, of their masses and
charge-changing weak transitions.  The relative strengths of these transitions
are parametrized by CKM matrix elements $V_{ij}$, where $V_{ud} \simeq V_{cs}
\simeq 0.975$, $V_{us} \simeq - V_{cd} \simeq 0.22$, $V_{cb} \simeq - V_{ts}
\simeq 0.04$, $|V_{ub}| \simeq 0.0036 \pm 0.0006$, and $|V_{td}| \simeq 0.009
\pm 0.002$ \cite{JRlat}.  The unitarity of this matrix implies that only four
of these parameters are independent.  Given their magnitudes, they are likely
to have relative phases which explain the observed CP violation in the kaon
system and imply observable effects for $B$ mesons \cite{JRlat}. 

\begin{figure}
\centerline{\epsfysize = 3 in \epsffile {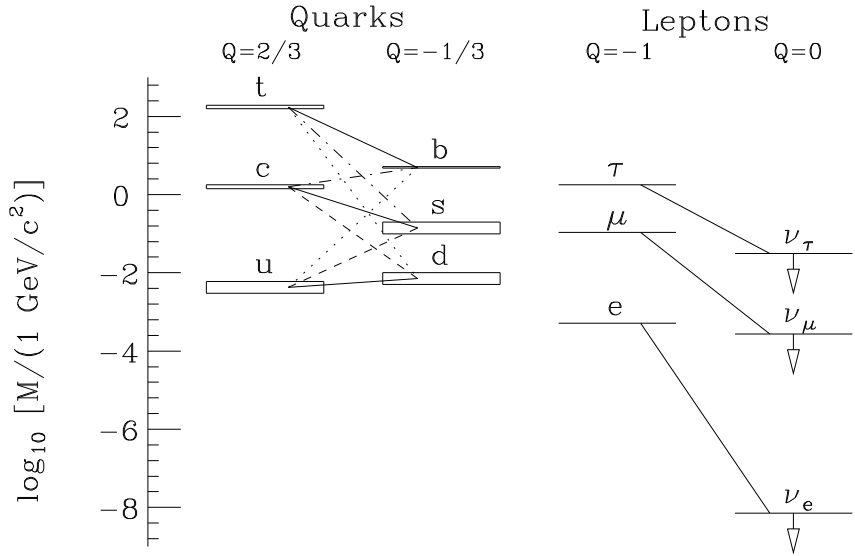}}
\caption{Patterns of charge-changing weak transitions among quarks and leptons.
Direct evidence for $\nu_\tau$ does not yet exist. The strongest inter-quark
transitions correspond to the solid lines, with dashed, dot-dashed, and dotted
lines corresponding to successively weaker transitions.} 
\end{figure}

The transition strengths in the CKM matrix are strongest for quarks in the same
family, weaker for quarks in neighboring families (numbering them in order of
increasing mass), and weakest for transitions between the first and third
families.  In this respect the pattern looks like that of electric dipole (E1)
transitions in atoms \cite{BS} or in quarkonium, where similar selection rules
have been noted \cite{GRED}.  The S-wave and P-wave $b \bar b$ levels are shown
in Fig.~2 \cite{JRAM}, together with diagonal lines denoting E1 transitions. 

\begin{figure}
\centerline{\epsfysize = 3 in \epsffile {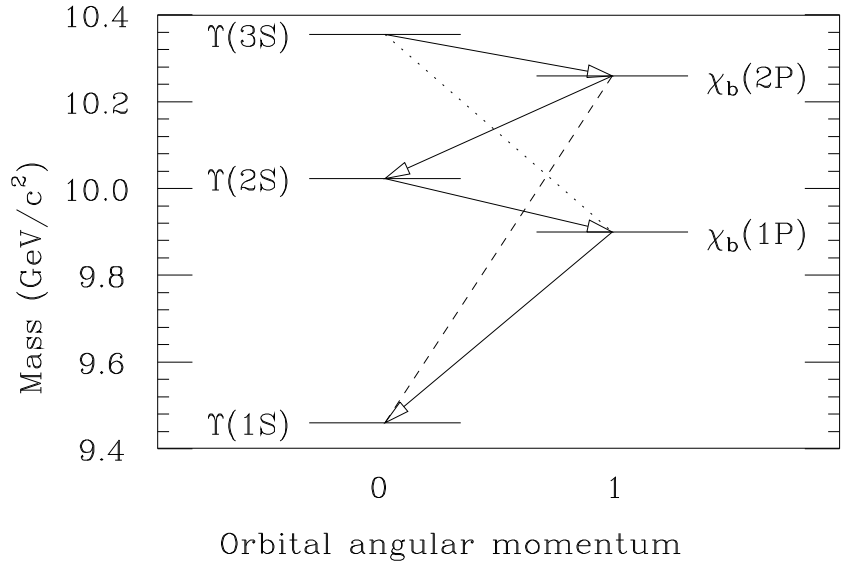}}
\caption{Patterns of electric dipole (E1) transitions between S-wave and
P-wave $b \bar b$ levels.  Solid lines denote strongest matrix elements;
dashed lines denote weaker matrix elements.}
\end{figure}

In atoms and quarkonia, transitions to ``nearest-neighbor'' states are favored,
as a result of greater overlaps between wave functions.  Thus, for example, the
dipole matrix element $\langle 3S|r|2P \rangle$ has a much larger magnitude
than $\langle 3S|r|1P \rangle$.  (Here we denote states by their radial, not
principal, quantum numbers.)  Hence the intensities of the photon lines in
$\Upsilon(3S) \to \gamma + \chi(2P)$ are much greater than those in
$\Upsilon(3S) \to \gamma + \chi(1P)$, far outweighing the advantage of
increased phase space in the latter decays.  In the case of transitions between
states of the three-dimensional harmonic oscillator, there is an exact
selection rule, since each cartesian component $x_i$ of the dipole operator is
a linear combination of creation and annihilation operators:  $x_i = (a_i
+ a^\dag_i)/\sqrt{2}$.  Thus E1 transitions in a three-dimensional harmonic
oscillator can only change the total excitation quantum number $N = N_x + N_y +
N_z$ by one unit. 

Any composite model of quarks and leptons must involve a mass scale $M_c$
larger than the experimental lower bound $\Lambda$ (of order one to several
TeV) describing deviations from pointlike behavior (as, for example, in the
production of jets at high transverse momenta in hadron-hadron collisions
\cite{jetlims}).  In principle $M_c$ could be as high as a grand unification
scale. In that case there is not much point to describing the quarks as
composites; there would then be no mass range in which the subunits would
exist as distinct entities.  However, we shall note in Sec.~4 that for Higgs
bosons a compositeness scale of one to several TeV is the highest scale for
which their composite nature is a useful concept.  If the composite natures of
Higgs bosons and fermions are related, as will be proposed in Sec.~5, there
will necessary be a large range of masses (from several TeV to a unification
scale) for which it makes sense to speak of subunits of quarks and leptons.

Other varieties of beyond-standard-model physics have mass scales intermediate
between the electroweak scale and the unification scale, though some are more
adept at concealing it \cite{Terning}.  The supersymmetry-breaking scale often
lies at such an intermediate value, while extended technicolor involves a large
mass scale (of order 100 TeV or greater) in order to avoid flavor-changing
neutral currents.  Thus the existence of a new mass scale above which quarks
and leptons can appear composite is not {\it per se} any more excess baggage
than accompanies other theories.

New opportunities for regarding fermions as composite \cite{Se} have arisen as
a result of the dualities between various Yang-Mills theories of gauge fields
and matter discovered by Seiberg and Witten \cite{SeWi}.  In these theories, if
the number of flavors and colors is chosen properly, there can exist zero-mass
fermionic bound states of other fermions, thereby automatically fulfilling the
't Hooft \cite{tH} condition.  Attempts (see, e.g, \cite{comps}) have been made
to construct realistic models based on these ideas.  A minimal goal for these
attempts would be to explain why we see only the three families of light
fermions illustrated in Fig.~1. The partial decay width of the $Z_0$ indicates
there are just three light weak-isodoublet neutrinos \cite{Zw}.  If there are
fermions heavier than those in Fig.~1, they must fall into different families
(at the very least, without light isodoublet neutrinos), perhaps in the manner
of the transition metals in the periodic table of the elements. 

A final reason to believe in compositeness of our present-day ``fundamental
units of matter'' is that it has happened previously at every scale at which we
have been able to look.  Starting with the subdivision of macroscopic matter
into molecules, molecules into atoms, atoms into electrons and nuclei, nuclei
into neutrons and protons (nucleons), and nucleons into quarks, structure
continues to change.  Typically such changes occur over length scales varying
anywhere from a factor of 10 to a factor of $10^5$ (the ratio between a Bohr
radius and the size of a proton).  Extrapolation from the electroweak scale of
100 GeV to a grand unification scale of $10^{16}$ GeV or the Planck scale of
$10^{19}$ GeV is a bold step, fraught with potential {\it hubris}.  It might be
correct.  The belief in compositeness of quarks and leptons is a conservative
``hedge'' against such an attitude.

\section{A special role for heavy top?}

The top quark, by virtue of its large mass and consquent large Higgs boson
coupling, could play a special role in electroweak symmetry breaking
\cite{specialt}.  A key question in such schemes is where one can expect new
physics beyond the standard electroweak picture.  If it begins just above the
top quark, the top quark is indeed special. It may be considerably less tightly
bound and more easily excited than other quarks.  On the other hand, if the
scale $M$ of new physics is at least 1 to 2 TeV, there is nothing particularly
special about the top. 

An example is provided by an Adler-Weisberger sum rule \cite{RRAW} for the
axial-vector coupling $g_A$ of the top quark.  One wishes to examine whether
deviations of $g_A$ from unity can provide information about the top
quark's structure \cite{PPZ}.  The characteristic scale for
deviations from $g_A = 1$ is $\Gamma_T/G_F M_T^3$, where $\Gamma_T$ and $M_T$
are the width and mass of the first excited state in the scattering between a
top quark and a longitudinal $W$ or $Z$.  For $M_T = {\cal O}(2$ TeV) and
$\Gamma_T$ characterized by the square of the weak SU(2) coupling constant 
times $M_T$, this deviation is at most a few percent.  This stands in sharp
contrast to the role of the $\Delta^{++}$ resonance in pion-nucleon scattering,
where the sum rule reads, in the SU(6) limit,
\beq
g_A^2 = \left( \frac{5}{3} \right)^2 = 1 + \left( \frac{4}{3} \right)^2~~~.
\eeq
The left-hand side is the nucleon pole contribution. On the right-hand side,
the first term refers to the equal-time commutator of two axial charges,
normalized by the current algebra, while the second term refers to the
$\Delta^{++}$ contribution.  The existence of a low-lying excited state of the
nucleon significantly affects its axial-vector coupling.  (An application of
the Adler-Weisberger sum rule to $g_A$ of the quark in the context of
large-$N_c$ QCD \cite{SWAX} is not devoted to the issue of quark compositeness
in the sense we are discussing here, but rather to models in which quarks and
pions are treated as separate degrees of freedom.) 

\section{Minimal technicolor:  QCD $\times$ 2650}

Imagine a world with zero-mass pions, coupling to the divergence of the
axial-vector current with a constant $f_\pi = 93$ MeV.  These then induce, via
the Higgs mechanism, a mass of about 30 MeV in the $W$, through mixing with its
longitudinal component \cite{STC}.  In order to induce a mass of 80 GeV in the
$W$, one needs an analogue of the zero-mass pion, but with a coupling to the
axial-vector divergence of $v = 2^{-1/4} G_F^{-1/2} = 246 {\rm~GeV} = (80
{\rm~GeV}/30 {\rm~MeV}) f_\pi$.  This boson $H^\pm$ is ``eaten'' by the $W^\pm$
and its neutral partner $H^0$ is ``eaten'' by the $Z^0$. 

Just as the pion is a quark-antiquark pair, bound to (nearly) zero mass by the
QCD interaction, imagine $H$ to be the state of a new fermion $F$ and
antifermion $\bar F$, bound with a ``technicolor'' interaction $v/f_\pi \simeq
2650$ times as strong as QCD.  The typical scale of hadrons (e.g., the mass of
the $\rho$ meson) is of order $2 \pi f_\pi$, as one can see, for example, from
a dynamical calculation based only on current algebra, unitarity, and crossing
symmetry \cite{BG}.  Thus, the typical scale of states characterized by the new
strong interaction should be about $2 \pi v$ or $v m_\rho/f_\pi = {\cal O}
(2$ TeV).  We expect vector $F \bar F$ states, for example, to have such
masses.  If we classify them according to ${\bf I} = {\bf I}_L + {\bf I}_R$,
they consist of an $I = 1$ state $\rho_T$ and an $I = 0$ state $\omega_T$.  The
suffix stands for TeV, their expected mass scale \cite{largeN,RRTC}.

The spinless composite $(U \bar U + D \bar D)/\sqrt{2}$ remains a particle in
the spectrum.  In analogy with the fate of the $\eta$ in QCD \cite{eta}, this
particle could acquire a mass characteristic of the electroweak symmetry
breaking scale.  

The axial current associated with the neutral pion is anomalous, accounting for
the decay of the pion to two photons \cite{ABJ} with an amplitude proprtional
to $Q_u^2 - Q_d^2$.  Here $Q_u = 2/3$ and $Q_d = -1/3$ are the electric charges
of the $u$ and $d$ quarks.  One wishes to avoid a corresponding anomaly in the
$Z - \gamma - \gamma$ coupling mediated by triangle graphs involving the
fermions $F$.  This is most easily satisfied by having the neutral Higgs boson
which is ``eaten'' by the $Z^0$ be of the form $(U \bar U - D \bar D)/
\sqrt{2}$, where $Q_U^2 - Q_D^2 = 0$ and $Q_U = Q_D + 1$, or $Q_U = - Q_D =
1/2$.  This ``minimal'' solution appears in the early technicolor
\cite{WTC,STC} and compositeness \cite{Akama} literature.  However, the
fermions $U,D$ are of no direct use in generating quark and lepton masses. 

\section{Fermion compositeness as a substitute for ``extended technicolor''}

In order to explain quark and lepton masses, one has to ``extend'' technicolor
in some manner \cite{ETC}, typically by introducing a proliferation of
techni-fermions (to implement anomaly cancellation) and new gauge bosons
connecting the ordinary and techni-fermions.  An alternative scheme can be
motivated in the following way \cite{RRTC}. 

Let us take the simple solution mentioned above for the technifermion charges: 
$Q_U = 1/2,~Q_D = -1/2$.\footnote{There exist tests for these charge
assignments based on the production and decay of the spin-1 composites $\rho_T
= (U \bar U - D \bar D)/\sqrt{2}$ and $\omega_T = (U \bar U + D \bar
D)/\sqrt{2}$ \cite{RRTC}.  These tests would have been possible at the planned
Superconducting Supercollider, but probably are out of reach of the
lower-energy Large Hadron Collider (LHC) at CERN.}  There is a simple
expression for the electric charge $Q$ of any quark and lepton \cite{chg}: 
\beq
Q = I_{3L} + I_{3R} + (B-L)/2~~~,
\eeq
where $I_{3L,R}$ are left-handed and right-handed isospin, $B$ is baryon
number, and $L$ is lepton number.  For all known quarks and leptons, $I_{3L} +
I_{3R} = \pm 1/2$.  Then let quarks and leptons contain a single subunit $U$ or
$D$, where this subunit carries the entire contribution to the electric charge
of $I_{3L} + I_{3R}$.  Moreover, let the chirality of the quark and lepton be
stricly linked to that of the $U$ or $D$.  Then to build quarks and leptons
incorporating a subunit $U$ or $D$ one merely needs another subunit
which (a) provides the contribution $(B-L)/2$ to the electric charge, and (b)
does not destroy the correlation between chiralities mentioned earlier.  The
simplest choice is a scalar $\bar S$:  $\bar S_{qi}$ with charge $1/6$ for a
quark and $\bar S_\ell$ with charge $-1/2$ for a lepton.  The index $i= 1,2,3$
denotes three colors; in accord with Pati and Salam's suggestion, lepton
number is the fourth color \cite{PS}.  We shall thus refer in what follows
to the extended (Pati-Salam) color group, including lepton number, as \suc.

The $F \bar S$ picture of quarks and leptons was proposed explicitly by
Greenberg and Sucher \cite{GS} in 1981, and is implicit in an earlier model
\cite{Akama}, in which $S$ itself is a composite of two fermions, one carrying
a horizontal (family) symmetry and the other carrying color or lepton number. 
It requires that one solve a dynamical problem to have {\it both} light $F \bar
F$ and $F \bar S$ states.  In any vectorlike theory for the force which binds
constituents into quarks and leptons, the chiral symmetry with which one hoped
to ensure small fermion masses \cite{tH}, instead of being realized in a
Wigner-Weyl sense, is expected to be spontaneously broken \cite{nogo}, so that
the lightest states are Nambu-Goldstone bosons rather than fermions.  The
chiral symmetry is then unavailable for protection of fermion masses.  One has
to postulate an effective interaction between fermions and scalars which
ensures that their lightest bound states are massless \cite{Suzuki}. 

Fermion masses in the present model would arise from a condensate $\langle U
\bar U + D \bar D \rangle \ne 0$ affecting quark and lepton masses in a manner
which depended on the bound state wave functions.  The exchange of scalars $S$
is seen to be a substitute for extended technicolor, as shown in Fig.~3.  In
order to learn about the $F \bar S$ Yukawa couplings to quarks or leptons, one
would have to solve the $F \bar S$ binding problem. 

\begin{figure}
\centerline{\epsfysize = 2 in \epsffile {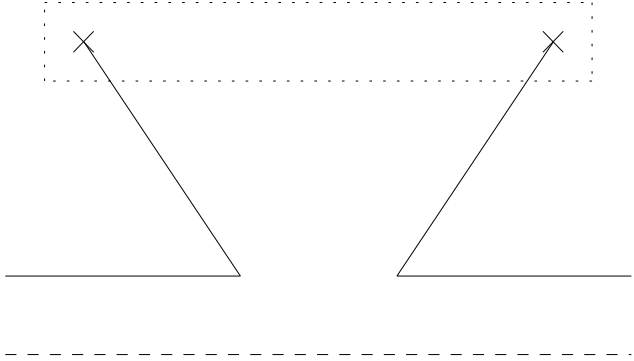}}
\caption{Diagram illustrating mass generation in composite model.  Solid lines
denote fermions $F$; dashed lines denote scalars $\bar S$.  The dotted
rectangle encloses a condensate $\langle U \bar U + D \bar D \rangle \ne 0$.}
\end{figure}

\section{Experimental signatures}

A typical new interaction in the $F \bar S$ model described above involves
a new contribution to the process $u \bar d \to e^+ \nu_e$ with a $U \bar D$
intermediate state.  (The $W^+$ also contributes to this process.)  Here
$u = U \bar S_q$, $\bar d = \bar D S_q$, $\nu_e = U \bar S_\ell$, $e^+ =
\bar D S_\ell$.  The $U \bar D$ intermediate state will typically have a mass
of order 2 TeV.  One may consider just the effect of a spin-1 state, coupling
via any combination of left- and right-handed couplings.  Effects will be
visible in production of lepton pairs at high transverse momenta ($p_T$) and in
high-energy $e^+ e^-$ annihilations \cite{BH,SR}, and in low-energy processes
such as $\mu \to e \gamma$ and $K$--$\bar K$ mixing \cite{GJJR}. 

As one example, one may consider the effect of an intermediate $\rho_T^+$ with
mass 2 TeV and Yukawa coupling $g_Y$ to a left-handed $u \bar d$ or $e^+ \nu_e$
pair \cite{SR}. In $p \bar p$ collisions at 1.8 TeV, when $\alpha_Y = g_Y^2/4
\pi = 0.1$, the cross section for production of a charged lepton above $p_T =
200$ GeV changes by about a factor of 2 up or down with respect to its nominal
value of about 4 fb, depending on whether the interference between the virtual
$\rho_T$ and the off-shell $W^+$ is constructive or destructive.  The
forward-backward asymmetry in $e^+ e^- \to b \bar b$ is affected by
direct-channel interference of a virtual $\rho^0_T$ with the virtual photon and
$Z$.  The effect, however, is only about $\pm 0.03$ for left-handed couplings
with $\alpha_Y = 0.1$, in an asymmetry expected to be a bit more than 0.50 at
$E_{\rm c.m.} = 200$ GeV \cite{SR}. 

In view of these modest effects, the direct search for excited states of the
Higgs boson at the LHC may be the best {\it entree} to a composite picture of
quarks and leptons.  If Higgs bosons are found to be composite, one may then
consider a similar possibility for quarks and leptons. 

\section{Effective supersymmetry in a composite model?}

In the $F \bar S$ model, the fermions $F$ consist of $U$, $D$, and their
corresponding antifermions, for a total of 8 degrees of freedom when spin is
included.  This is the same number of degrees of freedom possessed by the three
scalars $S_{qi}, S_\ell$ and the corresponding antiscalars.  One would be
tempted to see a manifestation of supersymmetry in this spectrum, except that
the charges of the fermions and scalars are not equal (as in many other
realizations of ``effective supersymmetry'' in condensed matter and nuclear
physics \cite{effSUSY}).  Does this really matter for the ultra-strong dynamics
responsible for $F \bar S$ binding?  If it does, could we have a phase
transition to equal charges for fermions and scalars at high energies?  For
example, by choosing the charges for the scalars $S$ to be $\pm 1/2$, one
obtains integrally-charged (Han-Nambu \cite{HN}) quarks, for which color and
electric charge do not commute with one another. 

It may be that supersymmetry provides the necessary ingredient to ensure light
composite fermions.  If so, we should either begin to see evidence of it in the
particle spectrum, or understand how a badly broken version of supersymmetry
can still be of use in this context. 

\section{Open questions in the ``F--S'' model}

If there is only a single family of scalars $\{S_{qi},S_\ell\}$, what degree of
freedom supplies the family structure?  Greenberg and Sucher \cite{GS} proposed
that different families corresponded to different radial excitations, with
orbital excitations lying significantly higher as a result of an effective
interaction close to a potential behaving as $r^{-2}$ near the origin.  One
would then have to understand CKM mixing in terms of overlaps of wave
functions. If, on the other hand, the scalars carry a family label, one loses
the motivation for a supersymmetric theory mentioned above, and one has to
introduce CKM mixing in an {\it ad hoc} manner. 

The pattern of quark and lepton masses and CKM matrix elements can be
described in terms of various hierarchical structures in mass matrices, of
which a recent and compelling example is given in Refs.~\cite{Ramond}.  Are
such hierarchies natural in composite models?

Do fermion masses arise from a condensate $\langle U \bar U + D \bar D \rangle
\ne 0$, as mentioned above?  What are the spectrum and wave functions of $F
\bar S$ bound states?  What about $S \bar S$ bound states?  Are they light? 
They could be a nuisance. If so, can we make them heavy? 

Are there other distinctive signatures of a generic $F \bar S$ model?  The
usual signatures of compositeness (see, e.g., \cite{sigs}) include: 
\begin{itemize}
\item Contact terms, e.g., new 4-fermion interactions
\item Excited states (at the compositeness scale)
\item Additional states which are light compared to the compositeness scale
\end{itemize}
and one would expect them to be present here as well.  In what follows we
construct a specific model for families based on a superstrong SU(4), which we
may call \sut.  This exercise illustrates both the pitfalls of explicit models
and their potential for predicting characteristic exotic states beyond the
usual quarks and leptons.

\section{Model based on a superstrong SU(4)}

Let both $F = U,D$ and a scalar $S$ be members of the fundamental ${\bf 4}_{\rm
T}$ of \sut.  Let the first family $f_1 \simeq u,~d,~\nu_e, e$ be $f_1 = F
\bar S$. Let the scalars in the second and third families be composites of $S$.
For example, if one chooses $S'$ to be a composite of three techni-antiquartets
$\bar S$ coupled up totally antisymmetrically to an \sut~quartet, it will
automatically be a ``flavor'' [\suc] quartet as well, i.e., it will correspond
to three quarks and a lepton, just like $S$. Thus the second family would have
the structure $f_2 = F \bar S' = FSSS$. 

The mixing of the first and second families can take place if there is an
operator in the Lagrangian transforming as $(SSSS)_{{\bf 1}_T}$. This operator
will also be an \suc~singlet if there are no relative angular momenta, by Bose
statistics. In turn, it can mix the second family with a third of the form $f_3
\equiv FS \bar S \bar S$, with the $S\bar{S} \bar{S}$ state in a $({\bf 4}_T^*,
{\bf 4}_T^*)$ representation. An operator transforming as $(S \bar{S})_{1_T}$
would mix $f_3$ with $f_1$.

The combination $S S \bar S$ contains {\it two} technicolor quartets rather
than one, so this model -- aside from having no known dynamical realization --
has serious shortcomings if we wish it to describe just the usual three
families of quarks and leptons.  The technicolor quartets, furthermore, are not
limited to be quartets of \suc.  Moreover, there seems to be no inherent
limitation to the number of additional $S \bar S$ pairs that can make up a
composite structure.  It may be that one has to consider instead the
possibility that the scalars themselves are not elementary and/or that they
carry family indices, as in Ref.~\cite{Akama}.  In any case, a detailed
calculation of $F \bar S$, $FSSS$, and $FS \bar S \bar S$ binding is required
to see if the model can yield sufficiently light quarks and leptons. One could
imagine performing such a calculation using lattice methods, for example, once
one learns how to treat chiral fermions on a lattice. 

The spinless $(S \bar S)_{{\bf 1}_T}$ states form exotic particles, consisting
of a neutral color octet, a color triplet with charges $2/3$, a color
antitriplet with charges $- 2/3$, and two neutral color singlets. The charged
colored scalars are leptoquarks, coupling to a charged lepton and a charge $\pm
1/3$ quark or a neutrino and a charge $\pm 2/3$ quark. They have to be quite
heavy $(\geq 100$ TeV) in order to suppress decays like $K_L \to \mu e$. This
is also true of the gauge bosons which become massive when \suc~breaks down to
SU(3)$_{\rm C} \times$ \ubl, if \sut~is a gauged symmetry.  If a condensate
$\langle ( S_1 \bar S_1 + S_2 \bar S_2 + S_3 \bar S_3 - 3 S_4 \bar S_4 ) /
\sqrt{12} \rangle \neq 0$ is responsible for breaking \sut~to $SU(3)_{\rm C}
\times$ \ubl, the leptoquark $S \bar{S}$ bosons could be absorbed into
longitudinal components of \suc / SU(3)$_{\rm C} \times$ \ubl~gauge bosons and
thus be removed from the low-lying spectrum. 

The absence of flavor-changing neutral currents at zero momentum transfer is
guaranteed by the orthogonality of wave functions of the mixtures of $F \bar S$,
$FSSS$, and $F S \bar S \bar S$ that make up the first three quark and
lepton families.  [A strong constraint on the compositeness scale may arise
from the apparent suppression of the decay $\mu \to e \gamma$ \cite{GJJR}, but
a detailed calculation is required for the present model.]  Since
charge-changing weak transitions involve the interchange $U \leftrightarrow D$,
any difference in residual interactions of a $U$ and $D$ with the scalars can
lead to non-trivial angles in the Cabibbo-Kobayashi-Maskawa matrix elements. 
This behavior has been illustrated recently in a different context \cite{RW}. 

It is not clear how one generates light neutrino masses in the present model.
The conventional picture involves a ``seesaw'' mechanism \cite{seesaw} with
large Majorana masses $M_M$ for right-handed neutrinos.  A suitable condensate
would seem to require at least two scalar fields $S_4$ in order to give the
required two units of lepton number violation, and a pair of fermions $F$ to
compensate the charge of the scalars.

Additional exotic technicolor-singlet combinations are possible.  For example,
one should be able to form $F \bar F$, $FFSS$, $FFFS$, and $FFFF$ states in
this model. The spin-zero $F \bar F$ states are the pseudo-Nambu-Goldstone
bosons which constitute the longitudinal components of the $W$ and $Z$, and one
massive state $(U \bar U + D \bar D)/\sqrt{2}$.  The spin-1 $F \bar F$ states
would be the techni-vector mesons $\rho_T$ and $\omega_T$ with predicted masses
of about 2 TeV. 

In the $FFSS$ states, the $SS$ subsystem must be in a ${\bf 6}_T$
(antisymmetric) representation.  Because of Bose statistics, the $SS_{{\bf
6}_T}$ pair must be antisymmetric in \suc, and so will consist of a color
triplet with charge --1/3 and a color antitriplet with charge 1/3. The $FF$ pair
is also in a ${\bf 6}_{\rm T}$ representation, and so must be {\it symmetric}
in its remaining degrees of freedom.  Thus, it consists of $I = J = 1$ and $I =
J = 0$ states, just as do the nonstrange quarks in the $\Sigma$ and $\Lambda$
hyperons.  The $FFSS$ states then consist of spin-1 color triplets with charges
2/3, --1/3, and --4/3 and antitriplets with charges 4/3, 1/3, and --2/3, and
spin-0 triplets with charge --1/3 and antitriplets with charge 1/3. Since the
$SSSS$ operator mentioned earlier can mix $FFSS$ with $FF \bar S \bar S$ (for
example), the $FFSS$ states have many features in common with diquarks. As a
result, their signatures in high-energy hadron collisions may not be very
distinctive. 

In the $FFFS$ states, the subsystem $(FFF)_{{\bf 4}_T^*}$ must be totally
symmetric in the product of its isospin $\times$ spin $[(I, J)]$ variables.
This is also true of the lowest baryonic quark-model states, which consist of
$N(1/2, 1/2)$ and $\Delta (3/2, 3/2)$ (for nonstrange states). Thus, we expect
the $FFF$ states to form an isospin doublet with charges $Q = \pm 1/2$ and spin
$1/2$, and an isospin quartet with charges $Q = (3/2, 1/2, - 1/2, - 3/2)$ and
spin $3/2$. The corresponding $FFFS$ states are then: 
\beq
J = 1/2 :~~~
\begin{array}{l}
{\rm Quarks ~ with} ~~ Q = (2/3,~- 1/3) \\
{\rm Leptons ~ with} ~~ Q = (0,~- 1) ~~~ , \\
\end{array}
\eeq
\beq
J = 3/2 :~~~
\begin{array}{l}
{\rm Quarks ~ with} ~~ Q = (5/3,~2/3,~- 1/3,~- 4/3) \\
{\rm Leptons ~ with} ~~ Q = (1,~0,~- 1,~- 2) ~~~ . \\
\end{array}
\eeq

The $FFFF$ states are found, by arguments similar to those presented for the
$FFF$ states, to consist of states with $(I,J) = (2,2), (1,1)$, and $(0,0)$. 
Orbital excitations of the lowest-lying states are expected. If experience with
ordinary hadrons is any guide, we expect them to lie about a TeV above the
corresponding $S$-wave states. 

\section{Families and spin configurations} 

An alternative picture of family structure is based on a simple example chosen
from quark-model baryon spectroscopy \cite{RW,PTP}.  The idea (in search of
a dynamics) is that when a down-type quark ($d,s,b$), containing a subunit such
as $D$, changes into an up-type quark ($u,c,t$), containing a subunit such as
$U$ by virtue of a weak transition, the changed interaction of the subunit
with the rest of the constitutents gives rise to a rotation of the eigenstates
of the dynamics in such a way that off-diagonal transitions are generated.

In the quark model one can see this behavior very transparently with baryons
consisting of three unequal-mass quarks (such as the charmed-strange baryons
$\Xi_c^{+,0}$).  The hyperfine interactions between the quarks, of the form
$\sigma_i \sigma_j/m_i m_j$, prevent the mass eigenstates from being diagonal
in the combined spins of any two of the quarks, though they are approximately
diagonal in the combined spin of the two lightest quarks \cite{GLR}.  In the
$\Xi_c^{+,0}$ the strange quark and the $u$ or $d$ quark are in an
approximately spin-0 configuration, with a small admixture of spin 1, while the
excited $\Xi_c^{'+,0}$ states, lying approximately 107 to 108 MeV higher
\cite{CLEOXic}, are mostly spin 1 in the two lightest quarks, with a small
admixture of spin 0. 

If one changes a $u$ quark to a $d$ quark, thereby changing $\Xi_c^+ = csu$
into $\Xi_c^0 = csd$, the mass eigenstates will rotate slightly with respect to
the basis states classified by the spin of the light-quark pair.  One thus
generates off-diagonal weak transitions between the excited and ground states.
The analogue of these transitions for composite quarks would be the
off-diagonal CKM matrix elements $V_{ij}~(i \ne j)$. 

Using only three fermions, one can only construct two states with spin 1/2. 
However, if one adds a unit of orbital angular momentum, one can construct
a model of three quark families, which, while artificial, illustrates the
principle \cite{RW}. 

\section{Summary}

We have reviewed some reasons for considering composite models of quarks and
leptons.  Foremost among these is the desire to understand the pattern of
masses and transitions in Fig.~1, which may signify a deeper level of
structure.  Analogies with previous experience, in atomic and nuclear physics,
may be misleading, but they strongly suggest that one may not need to solve all
outstanding dynamical problems before discerning the next layer of complexity.

We discussed one model in which the building blocks of quarks and leptons are
sets of fermions $F$ with charges $\pm 1/2$ and scalars $\bar S$ with charges
$1/6$ or $-1/2$.  Higgs bosons are spinless $F \bar F$ states, all but one of
which are absorbed into longitudinal components of $W^+, W^-$, and $Z$. In a
particular (and probably inadequate) example of this model based on a
superstrong SU(4), the three observed families are combinations of $F \bar S,
FSSS$, and $F S \bar S \bar S$. A rich spectrum of exotic states is predicted
to lie in the 1--3 TeV region, including spin-1 $F \bar F$ mesons and quarks
and leptons with unusual charges. 

Perhaps the pattern in Fig.~1 is incomplete.  Indeed, my favorite nonstandard
model involves not only these states but complete multiplets of the exceptional
group E$_6$, which includes isosinglet quarks and vector-like leptons
\cite{cmts,JRUCLA}.  So far I have not found a corresponding composite model,
\footnote{One could imagine a model based on constituents transforming
non-trivially under different parts of the SU(3)$_{\rm L} \times$ SU(3)$_{\rm
R} \times$ SU(3)$_{\rm C}$ subgroup of E$_6$, for example.}
but it would certainly be different from the one described above.  It may be
that just as in the case of the periodic table of the elements, it will be
{\it variations} in patterns which will provide the clue to underlying
structure.

\section*{Acknowledgments}

I wish to thank Chris Hill for the invitation to present these thoughts at the
workshop on New Strong Dynamics at Fermilab, and Rogerio Rosenfeld, Dave Soper,
Gerard Jungman, and Mihir Worah for collaboration on some of the topics
mentioned here.  At an early stage of these investigations useful discussions
were held in addition with G. Boyd, P. Freund, S. Gasiorowicz, T. Gherghetta,
J. Harvey, D. London, Y. Nambu, and S. Rudaz. This work was supported in part
the United States Department of Energy under Contract No. DE FG02 90ER40560. 

\def \ajp#1#2#3{Am.~J.~Phys.~{\bf#1}, #2 (#3)}
\def \apny#1#2#3{Ann.~Phys.~(N.Y.) {\bf#1}, #2 (#3)}
\def \app#1#2#3{Acta Phys.~Polonica {\bf#1}, #2 (#3)}
\def \arnps#1#2#3{Ann.~Rev.~Nucl.~Part.~Sci.~{\bf#1}, #2 (#3)}
\def \art{and references therein}
\def \bkly{{\it Proceedings of the 23rd International Conference on High Energy
Physics}, Berkeley, CA, July 16--23, 1986, edited by S. C. Loken (World
Scientific, Singapore, 1987)} 
\def \cmp#1#2#3{Commun.~Math.~Phys.~{\bf#1}, #2 (#3)}
\def \cmts#1#2#3{Comments on Nucl.~Part.~Phys.~{\bf#1}, #2 (#3)}
\def \corn93{{\it Lepton and Photon Interactions:  XVI International
Symposium, Ithaca, NY August 1993}, AIP Conference Proceedings No.~302,
ed.~by P. Drell and D. Rubin (AIP, New York, 1994)}
\def \cp89{{\it CP Violation,} edited by C. Jarlskog (World Scientific,
Singapore, 1989)}
\def \dpff{{\it The Fermilab Meeting -- DPF 92} (7th Meeting of the
American Physical Society Division of Particles and Fields), 10--14
November 1992, ed. by C. H. Albright \ite~(World Scientific, Singapore,
1993)}
\def \dpf94{DPF 94 Meeting, Albuquerque, NM, Aug.~2--6, 1994}
\def \efi{Enrico Fermi Institute Report No. EFI}
\def \el#1#2#3{Europhys.~Lett.~{\bf#1}, #2 (#3)}
\def \flg{{\it Proceedings of the 1979 International Symposium on Lepton
and Photon Interactions at High Energies,} Fermilab, August 23--29, 1979,
ed.~by T. B. W. Kirk and H. D. I. Abarbanel (Fermi National Accelerator
Laboratory, Batavia, IL, 1979)}
\def \hb87{{\it Proceeding of the 1987 International Symposium on Lepton
and Photon Interactions at High Energies,} Hamburg, 1987, ed.~by W. Bartel
and R. R\"uckl (Nucl. Phys. B, Proc. Suppl., vol. 3) (North-Holland,
Amsterdam, 1988)}
\def \ib{{\it ibid.}~}
\def \ibj#1#2#3{~{\bf#1}, #2 (#3)}
\def \ichep72{{\it Proceedings of the XVI International Conference on High
Energy Physics}, Chicago and Batavia, Illinois, Sept. 6--13, 1972,
edited by J. D. Jackson, A. Roberts, and R. Donaldson (Fermilab, Batavia,
IL, 1972)}
\def \ijmpa#1#2#3{Int.~J.~Mod.~Phys.~A {\bf#1}, #2 (#3)}
\def \ite{{\it et al.}}
\def \jmp#1#2#3{J.~Math.~Phys.~{\bf#1}, #2 (#3)}
\def \jpg#1#2#3{J.~Phys.~G {\bf#1}, #2 (#3)}
\def \ky85{{\it Proceedings of the International Symposium on Lepton and
Photon Interactions at High Energy,} Kyoto, Aug.~19-24, 1985, edited by M.
Konuma and K. Takahashi (Kyoto Univ., Kyoto, 1985)}
\def \Latt{{\it Lattice 98}, Proceedings, Boulder, CO, July 13--17, 1998}
\def \lkl87{{\it Selected Topics in Electroweak Interactions} (Proceedings
of the Second Lake Louise Institute on New Frontiers in Particle Physics,
15--21 February, 1987), edited by J. M. Cameron \ite~(World Scientific,
Singapore, 1987)}
\def \lnc#1#2#3{Lettere al Nuovo Cim.~{\bf#1}, #2 (#3)}
\def \lon{{\it Proceedings of the XVII International Conference on High Energy
Physics}, London, July 1974, edited by J. R. Smith (Rutherford Laboratory,
Chilton, England, 1974)}
\def \mpla#1#2#3{Mod.~Phys.~Lett.~A {\bf#1}, #2 (#3)}
\def \nc#1#2#3{Nuovo Cim.~{\bf#1}, #2 (#3)}
\def \nima#1#2#3{Nucl.~Instr.~Meth.~A {\bf#1}, #2 (#3)}
\def \np#1#2#3{Nucl.~Phys.~{\bf#1}, #2 (#3)}
\def \pisma#1#2#3#4{Pis'ma Zh.~Eksp.~Teor.~Fiz.~{\bf#1}, #2 (#3) [JETP
Lett. {\bf#1}, #4 (#3)]}
\def \pl#1#2#3{Phys.~Lett.~{\bf#1}, #2 (#3)}
\def \plb#1#2#3{Phys.~Lett.~B {\bf#1}, #2 (#3)}
\def \ppmsj#1#2#3{Proc.~Phys.-Math.~Soc.~Japan {\bf#1}, #2 (#3)}
\def \pr#1#2#3{Phys.~Rev.~{\bf#1}, #2 (#3)}
\def \pra#1#2#3{Phys.~Rev.~A {\bf#1}, #2 (#3)}
\def \prd#1#2#3{Phys.~Rev.~D {\bf#1}, #2 (#3)}
\def \prl#1#2#3{Phys.~Rev.~Lett.~{\bf#1}, #2 (#3)}
\def \prp#1#2#3{Phys.~Rep.~{\bf#1}, #2 (#3)}
\def \ptp#1#2#3{Prog.~Theor.~Phys.~{\bf#1}, #2 (#3)}
\def \rmp#1#2#3{Rev.~Mod.~Phys.~{\bf#1}, #2 (#3)}
\def \rp#1{~~~~~\ldots\ldots{\rm rp~}{#1}~~~~~}
\def \si90{25th International Conference on High Energy Physics, Singapore,
Aug. 2-8, 1990}
\def \slc87{{\it Proceedings of the Salt Lake City Meeting} (Division of
Particles and Fields, American Physical Society, Salt Lake City, Utah,
1987), ed.~by C. DeTar and J. S. Ball (World Scientific, Singapore, 1987)}
\def \slac89{{\it Proceedings of the XIVth International Symposium on
Lepton and Photon Interactions,} Stanford, California, 1989, edited by M.
Riordan (World Scientific, Singapore, 1990)}
\def \smass82{{\it Proceedings of the 1982 DPF Summer Study on Elementary
Particle Physics and Future Facilities}, Snowmass, Colorado, edited by R.
Donaldson, R. Gustafson, and F. Paige (World Scientific, Singapore, 1982)}
\def \smass90{{\it Research Directions for the Decade} (Proceedings of the
1990 Summer Study on High Energy Physics, June 25 -- July 13, Snowmass,
Colorado), edited by E. L. Berger (World Scientific, Singapore, 1992)}
\def \stone{{\it B Decays}, edited by S. Stone (World Scientific,
Singapore, 1994)}
\def \tasi90{{\it Testing the Standard Model} (Proceedings of the 1990
Theoretical Advanced Study Institute in Elementary Particle Physics,
Boulder, Colorado, 3--27 June, 1990), edited by M. Cveti\v{c} and P.
Langacker (World Scientific, Singapore, 1991)}
\def \Vanc{XXIX International Conference on High Energy Physics, Vancouver,
BC, Canada, July 23--29, 1998, Proceedings}
\def \yaf#1#2#3#4{Yad.~Fiz.~{\bf#1}, #2 (#3) [Sov.~J.~Nucl.~Phys.~{\bf #1},
#4 (#3)]}
\def \zhetf#1#2#3#4#5#6{Zh.~Eksp.~Teor.~Fiz.~{\bf #1}, #2 (#3) [Sov.~Phys.
- JETP {\bf #4}, #5 (#6)]}
\def \zp#1#2#3{Zeit.~Phys.~{\bf#1}, #2 (#3)}
\def \zpc#1#2#3{Zeit.~Phys.~C {\bf#1}, #2 (#3)}

\end{document}